\documentclass[sigconf,screen]{acmart}
\AtBeginDocument{%
  \providecommand\BibTeX{{%
    \normalfont B\kern-0.5em{\scshape i\kern-0.25em b}\kern-0.8em\TeX}}}

\setcopyright{acmlicensed}
\copyrightyear{2018}
\acmYear{2018}
\acmDOI{}
\usepackage{tabularx}
\usepackage{mdframed}

\acmConference[ESEM'24]{Make sure to enter the correct
  conference title from your rights confirmation email}{October 20--25,
  2018}{Barcelona, Spain}
\acmBooktitle{Proceedings of the ESEM 2024, October 20--25, 2024, Barcelona, Spain}

\begin{document}

\title{From Literature to Practice: Exploring Fairness Testing Tools for the Software Industry Adoption}

\author {Thanh Nguyen}
\affiliation{%
  \institution{University of Calgary}
  \city{Calgary}
  \state{AB}
  \country{Canada}
}
\email{thanh.nguyen2@ucalgary.ca}

\author{Maria Teresa Baldassarre}
\affiliation{%
  \institution{Università di Bari}
  \city{Bari}
  \state{BA}
  \country{Italy}
}
\email{mariateresa.baldassarre@uniba.it}

\author{Luiz Fernando de Lima}
\affiliation{%
  \institution{CESAR}
  \city{Recife}
  \state{PE}
  \country{Brazil}
}
\email{lffpl@cesar.org.br}

\author{Ronnie de Souza Santos}
\affiliation{%
 \institution{University of Calgary}
  \city{Calgary}
  \state{AB}
  \country{Canada}
}
\email{ronnie.desouzasantos@ucalgary.ca}

\begin{abstract}
\textbf{Context:} The increasing integration of artificial intelligence and machine learning into software systems has highlighted the critical importance of ensuring fairness in these technologies. Bias in software can lead to inequitable outcomes, making fairness testing essential. However, the current landscape of fairness testing tools remains underexplored, particularly regarding their practical applicability and usability for software development practitioners. \textbf{Goal:} This study aimed to evaluate the practical applicability of existing fairness testing tools for software development practitioners, assessing their usability, documentation, and overall effectiveness in real-world industry settings. \textbf{Method:} We identified 41 fairness testing tools from the literature and conducted a heuristic evaluation and documentary analysis of their installation processes, user interfaces, supporting documentation, and update frequencies. Technical analysis included assessing configurability for diverse datasets. The analysis focused on identifying strengths and deficiencies to determine their suitability for industry use. \textbf{Findings:} Our findings revealed that most fairness testing tools show significant deficiencies, particularly in user-friendliness, detailed documentation, and configurability. These limitations restrict their practical use in industry settings. The tools also lack regular updates and possess a narrow focus on specific datasets, which constrains their versatility and scalability. Despite some strengths, such as cost-effectiveness and compatibility with several environments, the overall landscape of fairness testing tools requires substantial improvements to meet industry needs. \textbf{Conclusion:} There is a pressing need to develop fairness testing tools that align more closely with industry requirements, offering enhanced usability, comprehensive documentation, and greater configurability to effectively support software development practitioners. \newline

\noindent \textbf{LAY ABSTRACT}. In today's world, we need to ensure that AI systems are fair and unbiased. Our study looked at tools designed to test the fairness of software to see if they are practical and easy for software developers to use. We found that while some tools are cost-effective and compatible with various programming environments, many are hard to use and lack detailed instructions. They also tend to focus on specific types of data, which limits their usefulness in real-world situations. Overall, current fairness testing tools need significant improvements to better support software developers in creating fair and equitable technology. We suggest that new tools should be user-friendly, well-documented, and flexible enough to handle different kinds of data, helping developers identify and fix biases early in the development process. This will lead to more trustworthy and fair software for everyone.

\end{abstract}

\keywords{software testing, software fairness, fairness testing}

\maketitle

\section{Introduction}
In recent years, the rapid growth of artificial intelligence has highlighted the importance of fairness in software engineering with respect to all phases of the development life-cycle. Software fairness is the ethical practice of ensuring that software systems, algorithms, and their outcomes are just, equitable, and unbiased across different groups of people, regardless of characteristics such as race, gender, ethnicity, or socioeconomic status~\cite{galhotra2017fairness}. In software engineering, this involves preventing discrimination, promoting inclusivity, and mitigating potential biases in the design, development, deployment, and usage of software applications and systems~\cite{brun2018software, zhang2021ignorance}.

Software fairness is essential across various societal aspects, especially as many organizations integrate machine learning into their processes, such as job interviewing, calculating credit scores, and assessing recidivism risk \cite{santos2024software}. The goal of software fairness is to eliminate biases so that when given a set of inputs that differ only on sensitive attributes (e.g., race, sex, age), the outcomes should be similar without targeting specific individuals or groups \cite{verma2018fairness, galhotra2017fairness}.  

Many researchers advocate treating software fairness as a first-class entity throughout the entire software engineering process \cite{chen2024fairness}. However, despite its importance, fairness testing remains under-explored and under-studied. For instance, it plays an important role in ensuring software systems meet fairness requirements by detecting and reporting bugs or faults resulting from system biases for future improvement \cite{chen2024fairness, zhang2020white, patel2022combinatorial}. The challenge lies in testing software fairness, as biases can arise at any stage of the development process. For example, biases can originate from early stages, such as training data and biased algorithms, to later stages, where user interactions feed data back into the system, perpetuating further biases \cite{chen2024fairness,santos2024software, santos2023perspective}.

As with any other type of testing in software development, the success of fairness testing is based on techniques, processes, and tools designed to support professionals in the industry \cite{raulamo2019practitioner, mustafa2009classification}. Testing tools are important because they provide the necessary infrastructure to identify, analyze, and fix issues within software systems. Therefore, given the limited number of studies focused on software fairness testing, this paper explores 41 software testing tools documented in a previous literature review \cite{chen2024fairness}. We used document analysis and heuristic evaluation to investigate how these tools could support software professionals in the industry who deal with fairness testing on a regular basis. Our goal is to answer the following research question: \textbf{RQ. \textit{How can existing software testing tools available in the literature support practitioners in the industry in conducting fairness testing?}}

This study makes key contributions to practitioners regarding software development and software fairness. First, by exploring fairness testing tools documented in the literature, we provide an overview of the current landscape of these tools. Second, through document analysis and heuristic evaluation, we explore several characteristics of these tools in supporting fairness testing in practice, including their strengths and potential limitations. Therefore, our study has the potential to equip practitioners with valuable insights into selecting and utilizing the appropriate tools for fairness testing, thereby enhancing their ability to identify, analyze, and mitigate biases in software systems. Additionally, our findings contribute to the broader understanding of how existing tools can be adapted or improved to better support fairness practices in the software industry. 


\section{Background} \label{sec:back}
With the advancement of technology, computers and their applications have become smaller, more portable, affordable, and integrated into many parts of our lives \cite{albusays2021diversity}. For example, previously, we cared about the sound quality when buying a speaker, but now we also consider whether it supports virtual assistants. As the importance of computers and their applications has grown, so has the emphasis on the quality of these applications \cite{pargaonkar2020review}.

Software testing is an essential element in the software development process, ensuring that products are bug-free before being released to the market. In this process, initially, software professionals used ad-hoc methods to manually detect errors, but today, advanced programs and testing tools have made the testing process easier and less time-consuming \cite{santos2020software, arzenvsek2014criteria}. Testing tools have become increasingly popular compared to manual testing because they can integrate smoothly into the automated software testing process, reduce labor, and eliminate human errors  \cite{alenzi2022survey, alferidah2020automated}.

Today, testing tools are widely available, varying in capabilities and features, which can make it challenging to choose the right tool for a specific testing purpose. When adequate tools are not used during testing activities, the susceptibility to errors in the software increases, often reducing the potential success of the software \cite{alenzi2022survey}. To select the appropriate tools, testing professionals rely on several characteristics and metrics, including compatibility with existing systems, ease of integration, the range of testing features provided (such as functional, performance, and security testing), user-friendliness, cost, and the level of support and documentation available \cite{sheth2015software, ekdahl2022mining, theivendran2023investigating, gamido2019comparative, fulcini2023review, da2023comparing}.

Previous research has explored and compared testing tools across various contexts. These studies have focused on defining metrics to evaluate these tools \cite{sheth2015software, ekdahl2022mining, theivendran2023investigating} and have drawn comparisons among tools in specific testing scenarios \cite{gamido2019comparative, fulcini2023review, da2023comparing}. However, the discussion on fairness testing is still in its early stages, and while several tools are being proposed, there are few studies focused on evaluating and comparing them based on defined features and metrics required by practitioners \cite{raulamo2019practitioner}. This gap highlights the need for systematic evaluations to ensure that the tools meet the practical needs of those working in the industry and effectively support fairness testing efforts.

\section{Method} \label{sec:method}
Our methodology builds upon established strategies from previous research that have explored and analyzed testing tools in various contexts \cite{raulamo2019practitioner, sheth2015software, ekdahl2022mining, theivendran2023investigating, gamido2019comparative, fulcini2023review, da2023comparing}. We incorporated general characteristics commonly required by testing practitioners and adapted standard metrics to develop a tailored set of criteria for analyzing fairness testing tools available in the literature.

\subsection{Tool Selection}
In the initial step, we searched for fairness testing tools that are open-source and publicly available. The fairness testing tools were collected from \cite{chen2024fairness} - Table \ref{tab:tools}. We began with 41 tools and then narrowed down the selection by evaluating each tool by following a set of inclusion and exclusion criteria.

\begin{table*}[ht]
\centering
\caption{Summary of Fairness Testing Tools}
\label{tab:tools}
\resizebox{\textwidth}{!}{%
\begin{tabular}{|l|l|l|l|}
\hline
\textbf{Tool} & \textbf{Type} & \textbf{Focus} & \textbf{Description} \\ \hline
FairTest & General ML & Model & Analyzing associations between outcomes and sensitive attributes \\ \hline
Themis & Classification (Tabular data) & Model & Black-box random discriminatory instance generation \\ \hline
Aequitas & Classification (Tabular data) & Model & Automated directed fairness testing \\ \hline
ExpGA & Classification (Tabular data, text) & Model & Explanation-guided fairness testing through genetic algorithm \\ \hline
fairCheck & Classification (Tabular data) & Model & Verification-based discriminatory instance generation \\ \hline
MLCheck & Classification (Tabular data) & Model & Property-driven testing of ML models \\ \hline
LTDD & Classification (Tabular data) & Data & Detecting which data features and which parts of them are biased \\ \hline
Fair-SMOTE & Classification (Tabular data) & Model & Detecting biased data labels and data distributions \\ \hline
FairMask & Classification (Tabular data) & Data & Extrapolation of correlations among data features that might cause bias \\ \hline
Fairway & Classification (Tabular data) & Data, ML program & Detecting biased data labels and optimal hyper-parameters for fairness \\ \hline
Parfait-ML & Classification (Tabular data) & ML program & Searching for hyper-parameters optimal to ML fairness \\ \hline
Fairea & Classification (Tabular data) & ML program, model & A unified benchmark for evaluating fairness repair algorithms \\ \hline
IBM AIF360 & Classification (Tabular data) & Data, ML program, model & Examining and mitigating bias in ML software \\ \hline
I\&D & Classification (Tabular data) & Model & Improving initial individual discriminatory instances generation \\ \hline
scikit-fairness & Classification (Tabular data) & Data, ML program, model & Examining and mitigating bias in ML software \\ \hline
LiFT & Classification (Tabular data) & Data, ML program, model & Examining and mitigating bias in ML software \\ \hline
FairVis & Classification (Tabular data) & Model & Visual analytics for discovering intersectional bias in ML software \\ \hline
BiasAmp & Image classifier & Model & Analyzing whether ML exacerbates bias from the training data \\ \hline
MAAT & Classification (Tabular data) & Data & Detecting selection bias and improving fairness-performance tradeoff \\ \hline
FairEnsembles & Classification (Tabular data) & ML program & Analyzing fairness and its composition in ensemble ML \\ \hline
FairRepair & Tree-based classification (Tabular data) & Model & Fairness testing and repair for tree-based models \\ \hline
SBFT & Regression (Tabular data) & Model & Search-based fairness testing for regression-based ML systems \\ \hline
ADF & DL-based classification (Tabular data) & Model & White-box fairness testing through adversarial sampling \\ \hline
EIDIG & DL-based classification (tabular data) & Model & White-box fairness testing through gradient search \\ \hline
NeuronFair & DL-based classification (tabular data, face images) & Model & Interpretable white-box fairness testing through biased neuron identification \\ \hline
DeepInspect & DL-based image classification & Model & Detecting class-based bias in image classification \\ \hline
CMA & Language models & Model & Detecting which parts of DNNs are responsible for unfairness \\ \hline
FairNeuron & DL-based classification (tabular data) & Model & Detecting neurons and data instances responsible for bias \\ \hline
RULER & DL-based classification (tabular data) & Model & Test input generation by discriminating sensitive and non-sensitive attributes \\ \hline
TestSGD & DL-based classification (Tabular data, text) & Model & Interpretable testing of DNNs against subtle group discrimination \\ \hline
DICE & DL-based classification (tabular data) & Model & Information-theoretic fairness testing and debugging of DNNs \\ \hline
ASTRAEA & NLP systems & Model & Grammar-based discriminatory instance generation for NLP systems \\ \hline
MT-NLP & NLP systems & Model & Metamorphic testing of fairness violation in NLP systems \\ \hline
BiasFinder & Sentiment analysis & Model & Metamorphic test generation to uncover bias of sentiment analysis systems \\ \hline
BiasRV & Sentiment analysis & Model & Uncovering biased sentiment predictions at runtime \\ \hline
NERGenderBias & Name entity recognition & Model & Measuring gender bias in named entity recognition \\ \hline
CheckList & NLP systems & Model & Behavioral testing (including fairness testing) of NLP models \\ \hline
DialogueFairness & Conversational AI & Model & Testing gender and linguistic (racial) bias in dialogue systems \\ \hline
BiasAsker & Conversational AI & Model & Fairness testing of conversational AI systems \\ \hline
REVISE & CV datasets & Data & Detecting object-, gender-, and geography-based bias in CV datasets \\ \hline
AequeVox & Speech recognition & Model & Comparing the robustness of speech recognition systems for different groups \\ \hline
\end{tabular}%
}
\end{table*}

\subsection{Inclusion and Exclusion Criteria}
Our inclusion and exclusion criteria were designed to ensure a practical evaluation of fairness testing tools, while also maximizing the number of tools analyzed. The criteria allowed us to simulate real-world conditions faced by testing professionals, focusing on the relevance and applicability of our findings. \\

\noindent \textbf{Inclusion Criteria.} We included tools that could be successfully installed and provided adequate instructions for use within a 2-hour evaluation timeframe. This criterion was chosen to simulate the behavior of a testing professional who is exploring a fairness testing tool available for their work in the industry. We also included tools that, although lacking comprehensive instructions, could be successfully installed and utilized within the 2-hour timeframe with the help of additional guidance from associated research papers. This approach allowed us to include tools that might otherwise be excluded due to minor documentation deficiencies, thereby exploring a broader range of available tools. \\

\noindent \textbf{Exclusion Criteria.} The exclusion criteria were applied to filter out tools that would not be practical for industry use. Tools were excluded if they could not be installed due to outdated programming languages, dependency conflicts, or broken source links. This ensured that only tools compatible with current technology standards were considered. We also excluded tools that lacked sufficient instructions for proper usage, as inadequate documentation could hinder the tool's usability and effectiveness in real-world projects.

By applying these criteria, we expected that our analysis focused on tools that were not only available but also feasible for use in a professional setting. This approach enabled us to explore the maximum number of tools possible, providing a comprehensive overview of the current landscape of fairness testing tools.

\subsection{Data Analysis}
We employed two data analysis methods on tools that passed our predefined inclusion and exclusion criteria: a heuristic evaluation focused on usability and documentary analysis. These methods ensured a practical evaluation of the selected fairness testing tools. \\

\noindent \textbf{Heuristic Evaluation.}
Heuristic evaluation is a usability inspection method used to assess a system against predefined characteristics \cite{nielsen1995conduct}. In our study, these characteristics were identified from previous studies as essential for evaluating testing tools \cite{raulamo2019practitioner, theivendran2023investigating, gamido2019comparative}. Our heuristic evaluation was designed to mirror the initial tool selection process performed by professional testers when considering tools for potential use in a project. 

Often, heuristic evaluations use practical guidelines to assess the usability of interfaces through walkthroughs and issue reporting. This approach is grounded in established rules.  In this paper, we evaluated several key characteristics of testing tools to ensure they are practical and effective for professionals in real-world settings \cite{raulamo2019practitioner, theivendran2023investigating, gamido2019comparative}. We looked at the ease of installation, including package requirements and compatibility. We explored the necessity for programming knowledge and script access, specifically, whether professionals need to modify the tool's scripts to run it effectively. The user-friendliness of the interface was evaluated for intuitiveness and accessibility. We checked the quality of documentation, particularly the presence and comprehensiveness of tutorial files like Readme.md instructions. The frequency of software updates was verified to ensure the tools remain current with technological advancements. Finally, we evaluated the versatility of each tool by examining its ability to handle various types of datasets and its adaptability to different scenarios. These characteristics are important for ensuring the tools are practical and effective for professionals in real-world settings. \\

\noindent \textbf{Document Analysis.}
Documentary analysis is a qualitative method used to review and interpret documents to explore and discuss a research problem. This method involves locating, interpreting, integrating, and drawing conclusions from valid documents such as guidelines, official reports, and academic papers \cite{fitzgerald2012documents, bowen2009document}. For this study, we conducted document analysis on the tool documentation, including user manuals, Readme files, and other instructional materials, guidelines associated with the tools, and the research papers in which the tools were initially proposed or evaluated. This approach enabled us to extract and synthesize relevant information, providing a comprehensive understanding of each tool's capabilities and limitations.

To support our documentary analysis, we employed thematic analysis \cite{cruzes2011recommended}, a method used to identify and analyze patterns (themes) within qualitative data. Thematic analysis is widely used in software engineering research, helping to identify cross-references among different data sources \cite{alhojailan2012thematic}. By systematically reviewing the documentation for each tool, we were able to highlight relevant characteristics and summarize our findings. This structured approach allowed us to gather detailed information, synthesize data from multiple sources, and draw conclusions about the tools' applicability and effectiveness in fairness testing, providing actionable insights for practitioners. \\

\noindent \textbf{Agreement Process.} Two researchers independently analyzed the 41 tools to ensure a thorough and unbiased evaluation. Each researcher conducted their assessment separately to avoid any influence from the other's findings. Following their independent analyses, a third researcher compiled and summarized the findings. Agreements between the two initial analyses were combined, and complementary findings were integrated to provide a comprehensive overview. Discrepancies or disagreements between the analyses were addressed in a consensus meeting, which could include the participation of a fourth author. This meeting facilitated a collaborative discussion to resolve differences and ensure a unified interpretation of the results. This process was straightforward, particularly because many tools could not be installed and, therefore, did not undergo heuristic evaluation and document analysis. This limitation reduced the volume of information requiring agreement among researchers.

\section{Findings} \label{sec:findings}
After applying our inclusion and exclusion criteria, only five tools identified in the literature met the requirements for our heuristic and documentary analysis: LTDD, which focuses on identifying and excluding unfair features in binary classification models; Fairea, which uses mutation to balance fairness and accuracy in binary classification models; Scikit-fairness, a versatile toolkit integrated into Python for evaluating fairness in both classification and regression tasks; FairRepair, designed to transform decision trees and random forests into fairer models while maintaining accuracy; and RULER, which improves fairness in deep neural networks through a phased training process. Each tool has been tested on various datasets, demonstrating different strengths and capabilities in enhancing fairness in machine learning models.

An essential factor for these tools passing our criteria was the success of their installation and ease of use within the allocated two-hour evaluation timeframe. Many tools available in the literature failed to meet our criteria due to insufficient documentation or instructions, leading to unsuccessful installations or requiring extensive modifications to operate effectively. This lack of detailed guidance and high modification requirements determined the exclusion of these tools from our further analysis, as testing professionals typically prioritize simplicity and ease of use when initially engaging with a tool. Below, our findings include both fundamental and specific characteristics of these tools, which are summarized in Table \ref{tab:tools_analysis}.

\subsection{Basic Requirement Assessment of Fairness Testing Tools}
By comparing five tools using the defined characteristics, we explored the basic characteristics of each one, highlighting the differences and commonalities among the tools in terms of installation, programming knowledge, and access to the code. For instance, among the five tools, only Scikit-fairness does not require downloading the source repository. For LTDD, Fairea, FairRepair, and RULER, users need to download the repository to install and use the tools. This involves setting up a virtual environment and installing necessary packages like NumPy and pandas.

Additionally, we noticed that all the tools necessitate some programming and machine learning knowledge, particularly in Python, to be used effectively. This requirement underscores the need for testing professionals to be familiar with coding and understanding machine learning principles to utilize these tools properly. Regarding access to the code, except for Scikit-fairness, which does not require users to access its source code, the other tools provide their source code for users. This means that users of LTDD, Fairea, FairRepair, and RULER can directly access and modify the underlying code if needed. However, to run Scikit-fairness, users must write their program to leverage its capabilities.

\begin{table*}[h!]
\caption{Comparison of Tools Based on Installation, Interface, Documentation, Updates, and Versatility}
\centering
\label{tab:tools_analysis}
\begin{tabular}{|l|p{2.7cm}|p{2.7cm}|p{2.7cm}|p{2.7cm}|p{2.7cm}|}
\hline
\textbf{Tool} & \textbf{Ease of Installation} & \textbf{User-Friendly} & \textbf{Documentation} & \textbf{Updats} & \textbf{Versatility} \\
\hline
\textbf{LTDD} & Easy; install repository, NumPy, Pandas, aif360 using virtual environment; minor syntax fixes needed & No, terminal interaction & Lacks detailed instructions in README; paper helpful & Last update: 2022, no recent commits & Single sensitive attribute; dedicated to specific datasets \\
\hline
\textbf{Fairea} & Easy; download repository, install modules from installation.md & No, terminal interaction; Jupyter Notebook & Good instructions in README and installation.md & Last update: 2021, no recent updates & Requires specific folder placement for datasets \\
\hline
\textbf{Scikit-fairness} & Very easy; pip install & Somewhat, interfaces for model comparison & Well-documented, dedicated web page & Regular updates, 15 versions, new soon & Versatile, supports various datasets \\
\hline
\textbf{FairRepair} & Easy; requires virtual environment for some packages & No, terminal interaction & Well-documented; README explains argument options & Last update: 2022, no recent updates & Scalability issues; longer repair times for large datasets \\
\hline
\textbf{RULER} & Easy; download repository, virtual environment for libraries & No, terminal interaction & README shows command lines but lacks detailed instructions & Introduced in 2022, no recent updates & Longer execution times; insufficient README explanation \\
\hline
\end{tabular}
\end{table*}

\subsection{Usability Characteristics and Limitations of Fairness Testing Tools}
We explored the general usability characteristics of each tool, focusing on ease of installation, the presence of a user-friendly interface, the quality and availability of tutorials or documentation, and the frequency of software updates. This analysis allowed us to understand the practical aspects and limitations of each tool and provide insights into its usability for testing professionals.

When assessing the process of installation, we observed that all tools are easy to install, but they require virtual environments for some external libraries such as NumPy and Pandas. We also found that to run Scikit-fairness, we need to write the whole program to assess fairness with unrestricted classification datasets, while others do not need to set up a new program but only offer certain datasets.

In terms of user-friendly interfaces, none of the tools were designed with user interfaces as part of their development. None of the tools were created to support professionals in designing and running fairness testing cases or testing fairness in general software. Instead, these tools are more focused on research or specific development contexts. This focus limits their practical application for professionals who require more accessible and streamlined tools to integrate fairness testing into their workflow.

For instructions or tutorials on how to use the tools, we conducted a review of the README files, installation guides, and associated research papers. Our findings indicate significant variability in the quality and comprehensiveness of the documentation provided for these tools. Specifically, LTDD lacks detailed information on how to execute the tool, making it challenging for users to get started. Similarly, RULER provides some details on usage but fails to offer comprehensive instructions, which can hinder effective utilization. In contrast, tools such as Scikit-fairness, Fairea, and FairRepair include well-structured and detailed documentation with clear instructions and examples that facilitate both installation and operation.

The frequency of updating the tools was another aspect of our heuristic evaluation. Regular updates are essential for maintaining the relevance and functionality of software tools. Upon examining the sources of each tool, we found that only Scikit-fairness receives regular updates that reflect ongoing improvements. This regular maintenance ensures that Scikit-fairness remains robust and adaptable to new challenges in fairness assessment. Conversely, other tools like LTDD, Fairea, FairRepair, and RULER have not been updated since the conclusion of their respective research studies. This lack of updates limits their long-term usability and effectiveness in real-world applications.

Finally, we assessed the versatility of each tool in handling various datasets. We observed that, with the exception of Scikit-fairness, all the tools are primarily designed for binary datasets, such as the commonly used Adult and COMPAS datasets. This limitation restricts their applicability in diverse real-world scenarios. Specifically, LTDD is not suitable for datasets containing multiple sensitive attributes, as it can only assess one sensitive attribute at a time. Fairea has a rigid requirement for dataset placement, necessitating that datasets be stored in a specific folder, which can be cumbersome. FairRepair's tree-based methodology demands significant time for training models, particularly with large datasets, making it less efficient. Similarly, RULER also generally requires a longer evaluation time, which can be a hindrance in time-sensitive applications.

\section{Discussions} 
In this study, we attempted to install and use 41 fairness testing tools available in the literature but were only successful with 5. These 5 tools were assessed based on their capabilities in supporting testing practitioners in industry tasks. Among them, Scikit-fairness emerged as the most effective tool.

Scikit-fairness offers several advantages: it is easy to install as a built-in package within the program, provides an instructive webpage to guide users, and receives frequent updates with refined versions. Additionally, while other tools are limited to specific binary classification datasets, Scikit-fairness can be used as a general tool to test most binary classification datasets. Regarding other tools, FairRepair performs well with small datasets due to its reliance on decision trees and random forests but needs enhancement to handle larger datasets efficiently. Similarly, RULER, which is based on a deep learning model, also needs optimization for faster execution times.

Looking at limitations, we observed that clearer instructions or tutorials that explain how to use the tools would significantly benefit practitioners. Only a few tools provide direct guidance from the source, while the majority often necessitates the reading of related research papers, which might not be efficient for practitioners. Regarding versatility, we noted that most tools are designed to handle only binary classification datasets, which limits their applicability. Scikit-fairness stands out as a more versatile option that is capable of handling a wider range of binary classification datasets. However, other tools have specific requirements or limitations that reduce their versatility. 

\subsection{What Practitioners Need from Software Fairness Tools}

According to the literature, practitioners require testing tools with several key characteristics, including applicability, compatibility, configurability, cost-effectiveness, cross-platform support, easy deployment, ease of use, expandability, maintenance of test cases, and test data, performance, popularity, and reporting features \cite{raulamo2019practitioner, gamido2019comparative, theivendran2023investigating}. These characteristics ensure that tools can be integrated seamlessly into various development workflows, cater to diverse datasets and user requirements, and support continuous improvement and adaptation to new challenges, which now include fairness requirements.

The five tools we analyzed exhibit some of these desirable characteristics but also show significant gaps. Most tools are open-source, making them cost-effective, but they often lack user-friendly interfaces and detailed documentation, which impedes ease of use. Compatibility is generally high with Python environments, yet the tools tend to be narrowly focused, handling only specific types of datasets and scenarios. Configurability and expandability are limited, as many tools do not offer sufficient options for customization or scaling to larger and more complex datasets. Additionally, regular updates and maintenance are lacking in many of these tools, which raises concerns about their long-term viability and support.

Testing tools that support daily activities in the industry are essential for elevating fairness to a first-class entity in software development. By incorporating fairness tools into standard development workflows, practitioners can identify and address bias issues early in the development cycle, preventing them from being embedded into deployed systems. This approach helps ensure that software products are fair and equitable, reducing the risk of negative societal impacts and enhancing the credibility and trustworthiness of the technology. Moreover, tools that are easy to use, well-documented, and integrated into existing systems empower practitioners to consistently apply fairness principles, making fairness an integral part of software development rather than an afterthought.

However, our study demonstrates that this is not the current scenario. Currently, tools are primarily research-focused, with limited applicability to real-world industry settings. They often lack user-friendly interfaces, detailed documentation, configurability, and regular updates, which limits their usability and effectiveness for practitioners. To better support industry needs, fairness testing tools should be developed to integrate seamlessly into development workflows and provide comprehensive reporting features to help identify and mitigate bias early in the development process, making fairness a fundamental aspect of software development and leading to more equitable and trustworthy technology solutions.

\subsection{Threats to Validity} 
Our analysis is inherently limited by the authors' specific expertise in software testing and software fairness. The first author is a junior practitioner working in the software development process for the government, including quality activities. The second author is a researcher specializing in empirical software engineering with a strong background in the human aspects of software development, including the perspectives of software practitioners such as testing professionals. The third author is a senior data scientist who is experienced in working on several machine learning projects. The fourth author has over eight years of professional experience in software testing, specializing in mobile testing, with over ten years of research in software quality and approximately three years focused on software fairness. This diverse combination of experience was leveraged to mitigate potential biases and provide a comprehensive evaluation of the tools.

Additionally, as a qualitative study that relies on heuristics based on previous studies and documentary analysis, our findings are subject to limitations inherent to these methods. The study focused exclusively on the official documentation of the tools, which may not capture all relevant information, and did not incorporate other data sources, such as the experiences of testers who used these tools. Nevertheless, this paper is designed for practitioners; hence, we chose a method that is straightforward and effective in producing actionable insights. Our goal was to inform practitioners about the available tools and highlight current needs for further research to enhance these tools in the context of software fairness.

\section{Conclusions} 
\label{sec:conclusions}
The growing importance of fairness in software systems, particularly those powered by artificial intelligence and machine learning, created a pressing need for effective fairness testing tools to be used in the software development process. Based on this problem, this study aimed to evaluate the potential of existing fairness testing tools in the literature to be used by testing practitioners and provide them with insights into the current landscape of tools while identifying areas where research was needed to better support fairness in software development.

While we identified 41 fairness testing tools in the literature, only 5 could be evaluated. These tools demonstrated some strengths, such as cost-effectiveness and compatibility with Python environments, but they also exhibited notable deficiencies. Many tools lacked user-friendly interfaces, detailed documentation, and configurability, which limited their applicability in real-world industry settings. Additionally, the tools were generally not maintained regularly and focused narrowly on specific datasets, which constrained their versatility and scalability. Among the tools analyzed, Scikit-fairness emerged as the most robust, offering regular updates and broader applicability, but even it had room for improvement, particularly in ease of use and documentation.

The insights from this study suggest several opportunities for future research. Currently, there is a clear need for the development of fairness testing tools that are more aligned with industry requirements, particularly in terms of user-friendliness, comprehensive work, and configurability. Future research could focus on creating tools that assist testing professionals in creating testing plans, implementing test cases, and identifying as well as offering robust reporting features to help them identify and mitigate bias early in the development process. Additionally, exploring ways to expand the scope of these tools to handle more complex and diverse datasets could significantly enhance their utility.  By addressing these gaps, researchers could contribute to making fairness a fundamental aspect of software development, ultimately leading to more equitable and trustworthy technology solutions.

\bibliographystyle{ACM-Reference-Format}
\bibliography{bib}

\end{document}